\documentstyle[12pt]{article}
\newtheorem{theorem}{THEOREM}
\newtheorem{lemma}{LEMMA}

\begin{document}

\begin{center}
{\Large \bf
Chaotic Spectra \\
of\\
Classically Integrable Systems}

\vspace{1truecm}
{\large \bf P. Crehan}
\\
{\it Dept. of Mathematical Physics, University College Dublin,\\
 Belfield,  Dublin 2, Ireland \\
PCREH89@OLLAMH.UCD.IE\\
Dept. of Mathematics, Faculty of Science, University of Kyoto,\\
Kyoto 606, Japan}
\end{center}

\begin{abstract}
We prove that any spectral sequence obeying a certain growth law is
the
quantum spectrum of an equivalence class of classically integrable
non-linear oscillators. This implies that exceptions to the Berry-
Tabor
rule for the distribution of quantum energy gaps of
classically integrable systems, are far more numerous than previously
believed. In particular we show that for each finite dimension $k$,
there
are an infinite number of classically integrable $k$-dimensional
non-linear oscillators
whose quantum spectrum reproduces the imaginary part of zeros on the
critical line of the Riemann zeta function.
\end{abstract}

\begin{center}
{\bf PCACS: Numbers 05.45, 03.65, 02.30.Dk, 02.30.Gp, 02.10.Lh}
\end{center}

\section{Introduction}
An important theme in the quantum theory of classically chaotic
systems
is the relationship between the qualitative behaviour of the classical
system and statistical properties of its quantum mechanical spectrum
\cite{eckhardt}. A much studied statistic
in this regard is the distribution of energy level spacings.
According to a result of Berry and Tabor the values of
the energy gaps of a generic integrable system are
Poisson distributed \cite{berry_tabor}. In general
this differs remarkably from the
statistical distribution of energy gaps in the case of
classically chaotic systems. These have been studied numerically
and can be described by Wigner, GOE or GUE rather than Poisson
statistics
\cite{eckhardt}.

An exception to the result of Berry and Tabor occurs in the case of
the harmonic oscillator. However Razavy investigated a family of
integrable
perturbations of the harmonic oscillator and found that
this departure from Poisson statistics is non-generic in the sense
that perturbing away from the harmonic oscillator, the energy gaps
quickly become Poisson distributed \cite{razavy}.
Another exception to the Berry--Tabor result was pointed
out by Casati, Chirikov and Guarneri, and occurs for the case of a
free particle in a rectangular well \cite{casati}.
Nevertheless Seligman and Verbaarshot observe that
for potential wells close to the rectangular well,
the distribution of energy gaps is
close to Poisson. They conclude that in this case also
departure from Poisson statistics is non-generic \cite{seligman}.
The general perception is that despite exceptional cases
such as the harmonic oscillator and the free particle in a rectangular
well, the statistics of quantum energy gaps of classically
integrable systems are universally described by the Poisson
distribution.

In this paper we demonstrate that given a spectral sequence obeying
a certain growth law, there exists an infinite family of classically
integrable Hamiltonians whose quantum spectrum coincides with this
sequence.
This shows that a wide range of exceptions to the Berry-Tabor rule
are possible and that any quantum system whose spectrum obeys such
a growth law, can be simulated by a family of classically integrable
non-linear oscillators. In particular we derive a result concerning
the hypothesis of Berry about the qualitative behaviour of an unknown
classical system whose quantum energy levels are given by the
imaginary part of the zeros on the critical line of the
Riemann-zeta function \cite{berry}. Berry argues that this classical
system should be chaotic. We prove that
this unknown classical system is not unique and need not be chaotic.
In fact we show that
an infinite number of classically integrable non-linear oscillators
are capable of reproducing these zeros when quantised.

\section{One dimensional case}
To justify our claims we require a number of theorems based on the
following lemma and its generalisation.

\begin{lemma}\label{lemma_1}
Given a sequence of complex numbers $\{\zeta_n: 0\leq n\in \it Z\}$
obeying
a growth law $ | \zeta_n | < \exp(a+b n)$ for some $a\in \Re$, $b\in
\Re^+$,
there exists an equivalence class $S$ of
entire functions of the complex plane such that for each
$s\in S$ we have $ s(n) = \zeta_n$.
\end{lemma}

\noindent We prove this by considering

\begin{eqnarray}
s(z) &=& \sum_{n=0}^{\infty}\zeta_n f_n (z)    \label{s(z)}
\end{eqnarray}

\noindent
where for $0\leq n \in \it Z$, $0<\epsilon \in \Re$ and $z\neq n$,

$$f_n (z) = \exp((z-n)(2\pi+b+\epsilon))\
{\sin(2\pi(z-n))\over 2\pi(z-n)}\ .$$

\noindent
Strictly speaking $f_n$ has a singularity at $z=n$. The
singularity is however removable and so defining
$f_n(n)=1$, $f_n$ becomes an entire function of the complex plane.
Eventually we will show that $s$ is well defined by the series
in eqn(\ref{s(z)}), and that it too is an entire function of the
complex plane.
Observing that $f_n(m) = \delta_{n m}$ for $0\leq n,m \in \it Z$,
the property $s(n)=\zeta_n$ follows from the definition of $f_n$ and
an
explicit evaluation of the sum at $z=n$.
This construction provides a single repesentative member of the
equivalence class $S$. The difference
between any two representatives is
an entire function of the complex
plane which vanishes on the non-negative integers. This
set is infinite and denoting it by $S_0$, the
equivalence class is given by $S=s+S_0$. All that now remains is to
show that $s$ is entire.

The domain $\{z\in C : |z| < \rho \in \Re\}$ will be denoted
$D_\rho$.
Using $A(D_\rho)$ to represent the analytic functions
on $D_\rho$, the entire functions are naturally denoted
$A(C)$. The Weierstrass M-test provides a criterion for when
the infinite sum of functions which are analytic on some domain
$D$, converges to a function which is analytic on $D$.
Specifically if $\{g_n\}$ is a sequence in $A(D_\rho)$, and
if there exists a sequence of positive real numbers $\{A_n\}$
with the property
$A_n \ge || g_n || = \sup_{z\in D_\rho} | g_n(z) |$,
such that $\sum_{n=0}^\infty A_n < \infty $, then $\sum_{n=0}^\infty
g_n$
converges to $g\in A(D_\rho)$.
To prove that $s$ is analytic on an arbitrary $D_\rho$, we
apply this test to the sequence defining $s$ in eqn(\ref{s(z)}).
Each term lies in $A(D_\rho)$ since each is
a constant multiple of an entire function.
Using the fact that $|sin(z)/z| \leq \exp(|z|)$ we have for $z\in
D_\rho$

\begin{eqnarray*}
| \zeta_n f_n(z) | & \leq & |\zeta_n|\
|\exp((z-n)(2\pi + b + \epsilon))|\ \left|
{\sin(2\pi (z-n))\over 2\pi(z-n)}\right| \\
&\leq & \exp(a+b n)\ \exp((|z|-n)(2\pi +b+\epsilon))\
\exp( 2\pi |z-n|)\\
&\leq & \exp(a+b n)\ \exp((|z|-n)(b+2\pi +\epsilon))\
\exp(2\pi (|z|+n))\\
&=&\exp(a+|z| (4\pi + b+\epsilon))\ \exp(-\epsilon n)\\
&\leq &\exp(a+\rho (4\pi+b + \epsilon))\ \exp(-\epsilon n).
\end{eqnarray*}

\noindent
To apply the Weierstrass M-test we take
$A_n = c_\rho \exp(-\epsilon n)$
where the constant $c_\rho = \exp(a+\rho(4\pi +b+\epsilon)$.
The sum $\sum_{n=0}^\infty A_n $ converges and so $s\in A(D_\rho)$.
Since $s\in A(D_\rho)$ for any $\rho\in \Re^+$ we have
$s\in A(C)$ and lemma \ref{lemma_1} is proven.

Although there
is a different $s$ for each valid choice of $a, b$ and $\epsilon$,
the corresponding equivalence class $S$ is independent of these
values. Lemma \ref{lemma_1} leads almost directly to the following
theorem.

\begin{theorem}\label{theorem_1}
Given a real sequence $\{E_n: 0\leq n\in\it  Z\}$ which obeys a
growth law $| E_n | \leq \exp(a+b n)$ for some $a, b\in \Re^+$,
there exists an equivalence class of classically integrable non-
linear
oscillators $H_C$ such that if $h^0$ is the Hamiltonian of the simple
harmonic oscillator, each $h_C \in H_C$ is of the form $h_C(h^0)$
and its quantum spectral sequence is given by $E_n$.
\end{theorem}

To see this construct $s$ as in lemma \ref{lemma_1} so that
$s(n)=E_n$ for $0<n\in Z$. The equivalence
class of classical Hamiltonians $H_C$ comes from replacing $z$ with
$h^0=(q^2+p^2)/2$ in each representative $s\in S$. As $h^0$ is
the Hamiltonian of the simple harmonic oscillator, each
$h_C\in H_C$ is the Hamiltonian of a classically
integrable non-linear oscillator.

Since $s\in A(C)$, each $h_C\in H_C$
can be identified with its Taylor expansion in $h^0$ which is
everywhere
convergent. This allows us to write
$h_C = \sum_{0}^\infty c_i\ (h^0)^i$ where each $c_i \in \Re$,
and to define a corresponding quantum Hamiltonian operator

\begin{eqnarray}
H_Q&=&\sum_{0}^\infty c_i N^i \label{ps_exp}
\end{eqnarray}

\noindent
where $N^i$ is the product of $i$-copies of the number operator.
With $\hbar =1$ $N$ is essentially the Hamiltonian of
the quantum harmonic oscillator. This procedure
quantises the classical Hamiltonian $h_C\in H_C$ and determines
an appropriate operator ordering.
The action of $H_Q$ on $L^2(x,dx)$ is well defined, the
eigenfunctions of $H_Q$ are the familiar harmonic oscillator
eigenfunctions, and the corresponding eigenvalues are simply $E_n$.

Although the coefficients of $H_Q$ in (\ref{ps_exp})
depend on the choice of $h_C\in H_C$, only values
of $h_C(z)$ when $z$ is a positive integer play a role in the
dynamics.
It does not matter which $h_C\in H_C$
we use to construct $H_Q$, the resulting quantum system
will be the same. $H_Q$ is a unique quantum Hamiltonian,
uniquely determined by the spectral sequence, and which we can
identify
with the equivalence class of classical Hamiltonians $H_C$.

This is a little surprising since it indicates that
the correspondence of classical to quantum systems
is not one-to-many as we might naievely have expected.
The usual correspondence between classical
observables $O_C(q,p)$, and quantum observables $O_Q(q,p,\hbar)$
is one-to-many in the sense that
for a given classical Hamiltonian $h_C(q,p)$, there is an infinite
number
of corresponding quantum Hamiltonians $h_Q(q,p,\hbar)$
determined by the property $h_Q(q,p,0)=h_C(q,p)$.
In our construction the redundancy due to different choices of
operator ordering does not arise. A different
redundancy however does arise. This is because there are
different classical systems $h_C$ which correspond to the
single quantum system $H_Q$. This is directly attributable
to the fact that many different continuous
functions interpolate a function whose values are specified only on
the integers. For a given value of Planc's constant
the differences between these classical systems occur on a
scale smaller than $\hbar$. For compact dynamical systems
such as those used in our construction, the correspondence between
classical and quantum systems is in fact many-to-many.

\section{Finite dimensional case}
It is straightforward to generalise the results of the previous
section to the $k$-dimensional case. A natural generalisation
of lemma \ref{lemma_1} is given by

\begin{lemma}\label{lemma_2}
Given a $k$-indexed sequence of complex numbers $\{\zeta_{n_1 \ldots
n_k}:
0\leq n_i\in \it Z$, $1\leq i \leq k\}$ obeying
a growth law $ | \zeta_{n_1 \ldots n_k} | < \exp(a+b_1 n_1 + \ldots
+b_k n_k)$
for $a\in \Re$, $b_i \in \Re^+$,
there exists an equivalence class $S$ of
entire analytic functions on $C^k$ such that $s(n_1,\ldots,n_k) =
\zeta_{n_1 \ldots n_k}$ for all $s\in S$.
\end{lemma}

This leads to a natural generalisation of theorem \ref{theorem_1}
as follows.

\begin{theorem}\label{theorem_2}
Given a $k$-indexed sequence of real numbers
$\{E_{n_1 \ldots n_k}: 0\leq n_i\in \it Z$,
$1\leq i\leq k\}$, which obeys the
growth law $| E_{n_1 \ldots n_k} | \leq
\exp(a+b_1 n_1 + \ldots + b_k n_k)$ for some $a\in \Re$, $b_i \in
\Re^+$,
there exists an equivalence class of classically integrable
$k$-dimensional non-linear oscillators $H_C$, where
if $h^0_i$ for $1\leq i \leq k$ are independent classical harmonic
oscillator Hamiltonians, each
$h_C \in H_C$ is of the form $h_C(h_1^0,\ldots ,h_k^0)$
and its quantum spectral sequence is $E_{n_1 \ldots n_k}$.
\end{theorem}

The proofs of lemma \ref{lemma_2} and theorem \ref{theorem_2}
are based on a consideration of

$$ s(z_1, \ldots, z_k) = \sum_{n_1 \ldots n_k =0}^{\infty}
\zeta_{n_1 \ldots n_k}\ f_{n_1}(z_1)\cdot \ldots \cdot f_{n_k}(z_k)$$

\noindent
where for $0<\epsilon_i \in \Re$ we have

$$ f_{n_i}(z_i) = \exp((z_i-n_i)(2\pi + b_i + \epsilon_i))
{\sin(2\pi (z_i - n_i))\over 2\pi (z_i-n_i)},$$

\noindent
along with the usual technical provision at each of
the removable singularities.
If $S_0$ is the set of entire functions of $C^k$
which vanish at the points
$(z_1, \ldots, z_k)=(n_1, \ldots ,n_k)$ for
$0\leq n_i \in \it Z$ and $1\leq i \leq k$, the equivalence classes
have the form $S = s + S_0$.
Once again they are independent of the explicit values of
$a, b_i$ and $\epsilon_i$ used to construct them.
The proofs proceed as before with only minor alterations and so
we omit the explicit details.

\section{Berry's hypothesis}
Comparing the spectral rigidity of quantum
systems which are classically integrable to those which are
classically chaotic, Berry  considered the spectrum of an
unknown dynamical system whose energy levels are given by the
imaginary parts of the zeros on the critical line of the Riemann zeta
function \cite{berry}. It had previously been conjectured by
Montgomery that the distribution of energy gaps of such a system
would be GUE \cite{montgomery}. Montgomery's conjecture was
supported numerically by the work of Odlyzko according to a report by
Bohigas and Giannoni \cite{bohigas}. Since GUE statistics are
normally associated with classically chaotic systems which do
not possess time reversal invariance, this suggested to Berry that
the corresponding unknown classical dynamical system
must be chaotic \cite{berry}.
Berry provided further support for this hypothesis through
theoretical
work based on a semi-classical consideration of the
rigidity of its spectral sequence \cite{berry}.

We will now show that although there may exist classically
chaotic systems whose quantum spectrum is given by
the imaginary parts of the non-trivial zeros of the Riemann zeta
function,
there also exists an infinite family of classical integrable systems
for which this is true.

To see that this is so it suffices to show that
the monotonic sequence $\zeta_n$ where $1/2+\imath \zeta_n$ is
the n-th nontrivial zero of the Riemann zeta function
satisfies the growth condition of theorem \ref{theorem_1}.
This assertion follows immediately from a classical theorem
about the distribution of zeros on the critical line due to
Hardy and Littlewood \cite{titchmarsh}. Their theorem states that
if $N(T)$ is the number of zeros of the Riemann zeta function
on the interval $[1/2, 1/2+\imath T]$, then there exists a constant
$a$ so that

$$ N(T) > a T.$$

\noindent This tells us that
if $1/2 + \imath \zeta_n$ is the position of the n-th zero, then

$$ \zeta_n < a^{-1} n.$$

\noindent
The sequence $\{\zeta_n: 0\leq n \in \it Z\}$ is therefore
exponentially bounded and satisfies the requirements of
theorem \ref{theorem_1}, which we apply to deduce the following.

\begin{theorem}
There exists an infinite family of classically
integrable non-linear oscillators whose quantum spectrum
is given by the imaginary part of the sequence of zeros
on the critical line of the Riemann zeta function.
\end{theorem}

It is possible to go even further by relabelling the
sequence with one index $\zeta_n$ as a sequence with
$k$ indices
$\zeta_{n_1 \ldots n_k}$ for any $1\leq k\in \it Z$,
such that
the growth condition of lemma \ref{lemma_2} is still satisfied.
There are many ways in which to do this and applying
theorem \ref{theorem_2} we deduce the following.

\begin{theorem}
For any finite dimension $k$, there exists an infinite family of
classically
integrable $k$-dimensional non-linear oscillators whose quantum
spectra
reproduce the imaginary part of the zeros on the critical line
of the Riemann zeta function.
\end{theorem}

\section{Conclusion}
Apart from an illustration of how the correspondence
between classical and quantum dynamical systems is
many-to-many rather than one-to-many,
our main conclusion is that exceptions to the rule
of Berry and Tabor regarding the distribution of energy gaps
in the spectrum of a classically integrable system,
are more numerous than the literature suggests.
In particular we show that contrary to the Berry hypothesis,
the unknown classical dynamical system whose chaotic quantum spectrum
is
given by the imaginary part of the
non-trivial zeros of the Riemann zeta function
is not unique and need not be chaotic. For a
given value of Planc's constant and for any finite
dimension of phase space, there exists an
infinite number of classically integrable non-linear
oscillators whose quantum spectrum simulates that
of Berry's unknown system.
We conclude that the Poisson distribution of energy gaps is
not a universal property of integrable systems, but of
a restricted class of systems for which the
approximations made by Berry and Tabor are valid.
It would be of considerable interest to
characterise more precisely the range of validity of
their approximations and consequently of their result
regarding the statistical distribution of quantum energy
gaps of classically integrable systems.

\section{Acknowledgements}
This work was started while an EU-STF fellow at the Dept. of
Mathematics in Kyoto University, and was finished at the Dept. of
Mathematical Physics at University College Dublin. We would like
to thank M. Jimbo and D. Judge for their hospitality, as well
as T. Shiota, P. Mellon, J. Lewis, E. Buffet, B. Eckhardt
and I. Guarneri for their comments and suggestions.

\end{document}